\documentclass[10pt, conference]{IEEEtran}
\IEEEoverridecommandlockouts

\def\BibTeX{{\rm B\kern-.05em{\sc i\kern-.025em b}\kern-.08em
    T\kern-.1667em\lower.7ex\hbox{E}\kern-.125emX}}
    
\usepackage{graphicx} 
\usepackage{xcolor}
\usepackage{url}
\usepackage{booktabs}
\usepackage{multirow}
\usepackage{tcolorbox}
\usepackage{enumitem}
\usepackage{threeparttable}
\usepackage{soul}
\usepackage{seqsplit} 
\usepackage{amsmath}
\usepackage{amssymb}
\usepackage{tabularx}
\usepackage{ragged2e}
\usepackage[numbers,sort&compress]{natbib}

\usepackage{hyperref}

  \usepackage{pifont}

\usepackage{framed}

  {\endMakeFramed}

\makeatletter
  \newcommand{\linebreakand}{%
    \end{@IEEEauthorhalign}                         
    \hfill\mbox{}\par
    \mbox{}\hfill\begin{@IEEEauthorhalign}
  }
  \makeatother

\begin{document}
\title{From Registry to Repository: How AI Agent Skills Are Written, Adapted, and Maintained}

\author{
    \IEEEauthorblockN{Haoyu Gao}
    \IEEEauthorblockA{
    The University of Melbourne\\
    Victoria, Australia\\
    haoyug1@student.unimelb.edu.au}
    \and
    \IEEEauthorblockN{Jai Lal Lulla}
    \IEEEauthorblockA{
    Singapore Management University\\
    Singapore\\
    jailal.l.2025@phdcs.smu.edu.sg}
    \and
    \IEEEauthorblockN{Hong Yi Lin}
    \IEEEauthorblockA{
    The University of Melbourne\\
    Victoria, Australia\\
    tom.lin@student.unimelb.edu.au}
    \linebreakand
    \IEEEauthorblockN{Sebastian Baltes}
    \IEEEauthorblockA{
     Heidelberg University\\
     Heidelberg, Germany\\
    sebastian.baltes@uni-heidelberg.de}
    \and
    \IEEEauthorblockN{Christoph Treude}
    \IEEEauthorblockA{
    Singapore Management University\\
    Singapore\\
    ctreude@smu.edu.sg}
    \and
    \IEEEauthorblockN{Mansooreh Zahedi}
    \IEEEauthorblockA{
    The University of Melbourne\\
    Victoria, Australia\\
    mansooreh.zahedi@unimelb.edu.au}
  }

\maketitle
\begin{abstract}

AI coding agents increasingly rely on \emph{skills}: structured context bundles, typically a \texttt{SKILL.md} file with a YAML header and Markdown body, loaded on demand for domain knowledge, workflows, and scripts. Public registries such as \texttt{skills.sh} now host tens of thousands of skills, making them an emerging unit of reuse in agent-based software engineering. Yet skills have largely been viewed as agent capabilities rather than software artefacts whose content and evolution shape agent behaviour. We present the first empirical study of AI agent skills as engineered artefacts that are authored, reused, customised and maintained, across public registries and personal-use repositories. We mined 18,463 skills from \texttt{skills.sh} and 23,199 personal-use skills from 5,876 GitHub repositories, identifying 3,709 reuse links. LLM-based classification into SWEBOK knowledge areas (KAs) shows Software Construction dominates alongside a long tail of specialised areas. A thematic analysis of 180 skills identifies six content categories.
Qualitative coding of 444 modifications reveals six themes, of which reworking operational specifications and adapting knowledge and resources are the primary target of change.
Our findings show that reuse is largely a one-time copy operation: most reused skills remain near-verbatim, 53\% are never modified after adoption, and subsequent local maintenance is overwhelmingly additive. Customisation primarily adapts skills to local environments, whereas evolution accretes new inline domain knowledge. Across both, a stable behavioural contract---how a skill interacts with users, monitors runtime state, and recovers from failures---remains almost untouched.
These results suggest maintenance effort should focus on project-specific bindings, and that registries and tool support should enable consolidating the domain knowledge skills re-author in isolation.

\end{abstract}

\section{Introduction}

Recent advances in large language models (LLMs) have given rise to coding agents that participate directly in software development, contributing to code generation~\cite{yu2024codereval}, code review~\cite{lin2025codereviewqa}, program repair~\cite{bouzenia2025repairagent}, test generation~\cite{schafer2023empirical}, and documentation~\cite{gao2026does}. To extend these agents into specialised tasks while keeping the context window manageable, the community has converged on the concept of skills: structured context bundles loaded on demand, supplying just-in-time domain knowledge, prescribed workflows, and optional companion scripts or reference documents. A skill typically consists of a \texttt{SKILL.md} file with a short YAML header and a Markdown body, packaged in a folder alongside optional assets. Skills reach an agent through two distinct channels. Some are published to centralised registries such as \texttt{skills.sh}~\cite{skills.sh}, which already hosts tens of thousands of skills, whilst others are authored by developers for the individual repositories and maintained alongside the source code as project-specific infrastructure. This mirrors the distinction between broadly reused third-party libraries and locally authored, project-specific components.

Emerging research on agent skills has concentrated on task coverage, safety, security, and performance. \citeauthor{ling2026agent}~\cite{ling2026agent} characterised skills' functional categories and identified a supply-demand imbalance. Security audits found that a substantial fraction of community-authored skills contain exploitable vulnerabilities, malicious payloads, or credential-leakage patterns~\cite{liu2026agent, chen2026credential}, while SkillBench measured whether skills improves agent task completion~\cite{li2026skillsbench}. Across these strands, the skill is examined as a capability layer rather than an engineered artefact that is authored, reused, and maintained. Yet, the engineering of skill content and evolution are equally critical. As reusable dependencies, skills can influence the quality, security, and maintainability of downstream systems, much as traditional dependencies shape software ecosystems~\cite{chowdhury2021untriviality, zimmermann2019small, decan2019empirical}. At the same time, as repository-resident documentation artefacts, skills risk becoming stale unless they evolve alongside the code they support ~\cite{gao2025adapting, aghajani2019software}. Skills also represent a novel form of software dependency. Unlike traditional packaged releases, they consist of editable natural language interleaved with executable code and routinely copied and adapted across projects. As a result, reused skills can diverge from their source and drift out of sync with the projects they support. Empirical understanding of this lifecycle is essential for standardising skill structures, informing authoring practices, and supporting long-term maintenance.

In this paper, we present the first empirical study of agent skills as engineered artefacts, spanning the registry and personal-use channels and tracing their reuse and maintenance. We mine 18,463 skills published to the centralised \texttt{skills.sh} registry and 23,199 personal-use skills from 5,876 GitHub repositories, recovering 3,709 reuse linkages that connect published skills to their locally adapted counterparts. Using SWEBOK knowledge areas (KAs) as a reference framework, we classify our corpus and obtain 29,450 SE skills in total to chart the software engineering knowledge they encode. We then characterise how skills are packaged and apply thematic analysis to 180 sampled \texttt{SKILL.md} documents for fine-grained content insights. Finally, we qualitatively code updates on 444 skills to derive change categories arising during customisation and evolution.


Our results identify Software Construction as the dominant KA, with the remaining KAs forming a long tail. Qualitative analysis of skill contents reveals six themes: 1) scoping and orchestrating, 2) running the execution lifecycle, 3) ensuring output quality, 4) governing agent conduct, 5) grounding domain knowledge, and 6) user/agent coordination.
Turning to how skills change once adopted, reuse is largely a one-time copy. Whether skills remain unchanged or evolve independently, they rarely incorporate subsequent updates of the centralised source. Our qualitative analysis of 444 updated skills surfaces six edit categories, namely: a) configuring skill metadata, b) reworking operational specifications, c) recalibrating behavioural constraints, d) adapting knowledge and resources, e) maintaining skill compatibility, and f) polishing presentation and wording. Reworking operational specifications and adapting knowledge and resources are the primary targets of maintenance, whilst rules governing agent conduct remain largely stable. Customisation re-grounds skills to their host projects, whereas evolution expands inline domain knowledge. The main contributions of this paper are as follows:

\begin{itemize}[leftmargin=*]
    \item A gold set and skills classification against SWEBOK KAs. 
    \item A taxonomy of \texttt{SKILL.md} content.  
    \item A taxonomy of maintenance behaviours for \texttt{SKILL.md} under the setting of customisation and evolution.
    \item Actionable implications to different stakeholders.
\end{itemize}

\section{Related Work}

\subsection{AI Agents and the Rise of Skills}


LLMs are now widely used in software development, from code generation to summarisation~\cite{chen2021codex, hou2024llms}. Initially, \emph{prompt engineering} steered them via handcrafted instructions and reasoning scaffolds to elicit desired behaviour~\cite{white2023prompt}. This static paradigm gave way to \emph{agents}, LLMs in perceive-reason-act loops that invoke external tools and iterate towards goals~\cite{yao2023react, schick2023toolformer}. Coding agents apply this loop directly to repositories, autonomously editing files to resolve issues~\cite{yang2024sweagent, jimenez2024swebench}, and have been surveyed as a distinct class of software engineering automation~\cite{liu2024agents}. As supplying capabilities ad hoc became a bottleneck, the Model Context Protocol (MCP) introduced an open standard exposing tools through a uniform interface~\cite{hou2025mcp, hasan2025mcp}. This trajectory marks a steady shift of intelligence from model weights into the dynamically supplied context.

This gave rise to declarative artefacts that supply context to agents. Repository-level files such as \texttt{AGENTS.md} and \texttt{CLAUDE.md} persist project conventions, build commands, and architectural notes, grounding it in a specific codebase. \citeauthor{chatlatanagulchai2025readmes}~\cite{chatlatanagulchai2025readmes} reported that these files mainly contain functional details rather than security or performance constraints, evolving through small edits like configuration code. A companion study notes Claude Code manifests share a shallow, single-heading structure~\cite{chatlatanagulchai2025manifests}, while an \texttt{AGENTS.md} file reduces execution time and token usage at comparable task completion~\cite{lulla2026agents}. \emph{Skills} extend these from passive descriptions to packaged, on-demand capabilities. A \texttt{SKILL.md} pairs preloaded YAML triggers with a Markdown body, optional companion scripts, and references disclosed only upon activation. Registries like \texttt{skills.sh} have indexed tens of thousands of skills, establishing a new unit of reuse in agent-based development. Nascent research examines skills mainly as capability layers. \citeauthor{ling2026agent}~\cite{ling2026agent} charted functional categories and surfaced supply-demand imbalances, security audits revealed vulnerabilities and credential-leakage patterns~\cite{liu2026agent, chen2026credential}, and SkillBench evaluated their impact on task completion~\cite{li2026skillsbench}. Yet, the skill remains unexamined as an \emph{engineered artefact}, leaving its authoring, structuring, reuse, and maintenance as surrounding code evolves largely unexplored. 

\subsection{Reuse and Evolution of Software Artefacts}

Contemporary software is assembled rather than written from scratch, relying on third-party components from registries such as npm and PyPI~\cite{decan2019empirical, wang2020thirdparty}. This reuse boosts productivity but introduces dependency chains underestimated even for trivial packages~\cite{chowdhury2021untriviality}, turning registries into shared attack surfaces. A single npm package can reach over 100{,}000 dependents, so a compromised maintainer account can infect much of the ecosystem~\cite{zimmermann2019small}, while malicious supply-chain packages have been dissected from real-world cases~\cite{ohm2020backstabber} and systematised into attack taxonomies~\cite{ladisa2023sok}. A recent strand extends this to the \emph{AI supply chain}, where pre-trained models reused from hubs such as Hugging Face exhibit provenance, turnover, and documentation-quality dynamics distinct from conventional packages~\cite{jiang2023ptmreuse, jiang2024peatmoss}.

Reused artefacts, however, are never finished. Lehman's laws cast this as continuing change, where real-world systems must continually adapt to prevent deteriorating utility and unchecked structural decay~\cite{lehman1980programs}. Documentation is particularly susceptible because it describes a moving target. Prose and code seldom evolve in lockstep, accumulating inconsistencies~\cite{wen2019codecomment} and making out-of-date content a prevalent issue in documentation taxonomies~\cite{aghajani2019software}. The consequences are concrete, as installation instructions silently break amid underlying ecosystem shifts~\cite{gao2025adapting} and once-functional tutorials become unexecutable~\cite{mirhosseini2020docable}. A complementary research track therefore seeks to characterise and repair documentation, ranging from categorising README content~\cite{prana2019readme} to augmenting API documentation with mined caveats and crowd knowledge~\cite{treude2016augmenting, li2018improving}.

\begin{figure*}[ht!]
    \centering
    \includegraphics[width=0.65\linewidth]{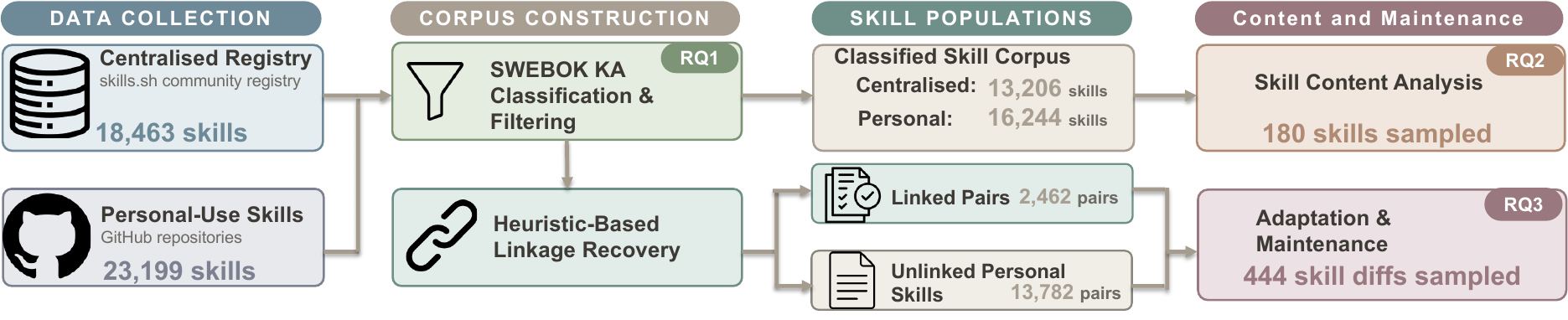}
    \caption{Study design: we collect centralised and personal-use skills and recover their reuse links, then answer RQ1 (SWEBOK knowledge-area classification), RQ2 (packaging and content), and RQ3 (customisation and evolution).}
    \label{fig:research-diagram}
\end{figure*}

Skills inherit both pressures. As an emerging supply chain, they span the third-party and personal-use forms recognised in reuse research. Published registry skills are the \emph{third-party} kind: canonical procedures whose qualities and flaws propagate to every adopter. Just as trivial packages introduce dependency chains~\cite{chowdhury2021untriviality}, flawed skills propagate vulnerabilities~\cite{liu2026agent} or leaked credentials~\cite{chen2026credential}. Personal-use skills, by contrast, are \emph{in-house}, encoding undistributed but equally consequential project-specific knowledge. Unlike compiled, version-pinned releases, skills interleave natural language with code, an unusually mutable dependency susceptible to customisation upon adoption and silent drift during in-place maintenance. The stakes also differ from documentation. A human reading an out-of-date README can identify and bypass the discrepancy~\cite{mirhosseini2020docable}, whereas an agent consumes a \texttt{SKILL.md} and acts on its environment, so stale guidance is executed rather than read. Our study traces both, reconstructing links between registry skills and their personal-use copies to characterise how developers customise and maintain this reuse.

\section{Methodology}

We structured our research into three RQs.

\noindent \textbf{RQ1}: What software engineering KAs do skills cover?

\noindent \textbf{RQ2}: What are the characteristics of AI agent skills?

\textbf{RQ2.1}: How are they written and packaged? \textbf{RQ2.2}: What are their contents?


\noindent \textbf{RQ3}: How do open-source software repositories manage and adapt AI agent skills over time?

\textbf{RQ3.1}: To what extent are \texttt{skills.sh} skills modified when transferred into repositories? \textbf{RQ3.2}: What changes do developers make to skills after their initial addition?

These three RQs are designed to capture a comprehensive view of the newly emerged artefact of skills. As coding agents play an increasingly important role, our RQ1 investigates what software engineering knowledge areas are covered by AI agents to assist during development, establishing a foundational understanding of the nature of these skills. Following this, RQ2 progressively examines the bundled content within skill artefacts across various software development contexts. Finally, RQ3 systematically explores how these skills are adapted and maintained over time. As newly emerged software artefacts, their usage and maintenance patterns differ from those of traditional code reuse and self-constructed components. Figure~\ref{fig:research-diagram} illustrates the overall workflow of our study.

\textbf{Terminology}. Throughout our methodology and dataset, we define three skill paradigms. \emph{Centralised skills} are published to the skills.sh registry. \emph{Personal-use} skills are collected from GitHub, residing alongside project source code and not in the registry. Among these, we classify a skill as \emph{linked} if our heuristic matches it to a centralised skill (representing registry reuse), and \emph{locally authored} or \emph{unlinked} otherwise.


\subsection{Dataset Collection and Construction}

\ul{\textbf{Centralised Skills Collection}}: We selected skills.sh as our centralised source, as it is the primary community registry for AI agent skills. All skills hosted there are fully open source and traceable to GitHub, and the platform only registers skills that developers explicitly add via the ``\textit{npx skills add}'' command. By contrast, alternative platforms such as SkillMP periodically scrape skills from open-source repositories and rely on project stars to estimate popularity, a metric that may not reflect a skill's true adoption.

To construct the dataset, we queried skills.sh for all skills with at least 20 installs as of 4 March 2026, yielding 19,830 skills across 2,419 GitHub repositories. We then retrieved the raw files via the provided links to their original repositories, collecting each \texttt{SKILL.md} together with its co-existing assets. Of the 19,830 indexed skills, 18,463 from 2,338 repositories were successfully retrieved. The remainder were inaccessible because their repositories had been deleted or made private.

\ul{\textbf{Personal-Use Skills Collection}}: We queried the GitHub Search API over the same period with the filter \texttt{filename:SKILL.md}, sharding queries by file size to bypass the result limit. To avoid duplication, we excluded skills from repositories already in our centralised collection, and omitted forks, archived repositories, and those with fewer than 10 stars. This yielded 110,090 skills across 6,876 repositories. We further dropped repositories whose name or description contained collection-indicator keywords (e.g., awesome, collection, tutorial, course, registry), so repositories represent real projects rather than aggregators, leaving 6,613 with 76,921 skills. The per-repository skill count remained skewed, indicating the keyword filter missed some aggregators. Manual inspection confirmed this tail is mostly skill libraries and marketplaces (e.g., an 8,187-skill archive), so removing them prunes aggregators, not genuine projects. We therefore discarded repositories whose skill count exceeds Tukey's fence (18 skills in our distribution)~\cite{tukey1993exploratory}, yielding a final 5,876 repositories with 23,199 skills.

\ul{\textbf{Reuse-linkage Construction}}: We then linked the two populations. The key enabler is the YAML \texttt{name:} field that agents use to progressively load skills. When multiple centralised skills shared a name, we disambiguated by content similarity between the personal-use \texttt{SKILL.md} body and each candidate, using the longest-common-subsequence ratio:
$\text{sim}(A, B) = \frac{2 \times |LCS_{chars}(A, B)|}{|A| + |B|}$.
This captures verbatim reuse. Specifically, a near-identical copy scores close to 1, while an independently written skill of the same name scores near 0.

A personal-use skill with no name match, or whose highest similarity fell below 0.1, was classified as \textit{unlinked}. Among multiple name matches, the candidate with the highest similarity above 0.1 was taken as the linked pair. We treated a pair as genuine reuse only when it showed textual or structural correspondence beyond a shared subject, such as verbatim passages, identical section structures, or matching idiosyncratic details like configuration values. To validate the heuristic, two authors independently audited the boundary bucket $[0.1, 0.3)$, where false links are most likely. They agreed on 90\% of 30 sampled pairs and judged 28 (93\%) to be genuine reuse. We assume this judgement grows more reliable as similarity increases, though the scheme loses recall for heavily adapted or renamed skills.
Conversely, to confirm the threshold does not discard genuine links, we sampled 30 name-matched pairs below 0.1. All 30 were independent. Together these validate 0.1 as the linkage threshold.
Finally, this produced 3,709 linked pairs spanning 1,673 centralised skills (8.5\% of the centralised collection), with 70.3\% of pairs at similarity $\geq$ 0.99, indicating near-verbatim copying. On the personal-use side, 16.0\% of skills trace back to a centralised registration and 76.0\% are unlinked, the rest having no YAML field and so left unlinked.

\subsection{RQ1 Methodology}
\label{sec:rq1-method}

To obtain a comprehensive collection of software engineering knowledge, we consult SWEBOK v4~\cite{SWEBOK}, a consensus-driven reference reflecting the current state of software engineering theory and practice. It comprises 18 KAs, each specifying a distinct domain with its principles and practices. Published by the IEEE Computer Society and recognised by ISO/IEC, SWEBOK is an authoritative reference for the scope of software engineering knowledge.
To classify skills into KAs with LLMs, we retrieved the 18 primary KAs and their immediate sub-topics and formalised them into a structured prompt, guiding the models to assign each skill, from its name and description, to one of the 18 KAs or a non-SE category. For reproducibility, we set the temperature to 0 and release the prompts and other settings in our replication package (Section~\ref{sec:replication}).
Following \citeauthor{ahmed2025can}~\cite{ahmed2025can}, we employed two open-source LLMs, Qwen3.6-27b and Gemma-4-31b-it, to classify a random sample of 400 skills.
Their initial Cohen's $\kappa$ was 0.61, indicating substantial agreement~\cite{mchugh2012interrater}. We then examined disagreements and enriched the prompt with finer-grained sub-subtopic examples per KA, five to six each chosen through discussion, raising agreement to 0.79. This calibration ensured the models were accurate enough to sample a representative set of skills per KA for subsequent manual evaluation.

To build a gold standard for the 19 categories, we drew a stratified sample of 15 skills per category (285 in total), running Qwen3.6-27b over a larger pool to ensure at least 15 predicted skills each. Because the sample is balanced by predicted label rather than corpus proportion, it characterises category-specific annotation quality rather than corpus-level accuracy. Two authors independently annotated the set, reaching a Cohen's $\kappa$ of 0.68 initially and 0.85 after a second round resolving disagreements, with remaining conflicts reconciled into the gold standard. The resulting set ranges from 20 skills (Software Construction) to 11 (Software Architecture, Software Testing, Software Engineering Management, and Software Engineering Models and Methods), the departure from a uniform 15 arising because skills were sampled by the LLM's label, which the gold annotations sometimes reassigned to other KAs. Against this gold standard, Qwen3.6-27b yielded a Cohen's $\kappa$ of 0.82, so we applied it to all collected skills.


\subsection{RQ2 Methodology}
We investigated each skill as a package, analysing its \texttt{SKILL.md} alongside any optional assets in the same directory. We examined structural and textual features such as document length, sectioning, and asset prevalence. We additionally drew on a public skill specification supported by Anthropic~\cite{skillspec}, which prescribes the standard directory structure, required and optional frontmatter fields, subdirectory layouts, and body-length recommendations. From it we extracted 17 statically inspectable provisions, setting aside soft clauses requiring semantic judgement (e.g., ``\emph{the description should describe what the skill does and when to use it}''). These comprised 7 mandatory and 10 optional provisions, against which we computed compliance rates across both populations.

For the thematic analysis, we performed stratified sampling of skill documents across KAs~\cite{baltes2022sampling}. Given the highly skewed distribution over KAs, we sampled 10 per KA, yielding 180 documents. Each document was split into paragraphs, then merged into coherent sections as the unit of analysis. Two authors first jointly coded a shared set of 20 skills to establish an initial code set, allowing multiple codes per section. After two rounds (40 skills) agreeing on code granularity and borderline cases, they coded the remaining skills independently and cross-checked each other's assignments, revising to reconcile disagreements. Throughout, following an established thematic analysis approach~\cite{cruzes2011recommended}, we compared emerging codes within and across skills, grouping them into subthemes and overarching themes concurrently. The categorisation was finally reviewed and agreed by the whole team.

\subsection{RQ3 Methodology}



RQ3 examines how skills change after entering a repository, along two axes. (1) \textit{Customisation} captures how reused skills are adapted. For each linked pair, we compute the immediate diff between the personal copy and the centralised snapshot at adoption version recovered from upstream history. (2) \textit{Evolution} captures how locally authored skills are maintained over time, tracing the \texttt{SKILL.md} commit history of each unlinked skill. The evolution diff compares only a skill's initial and final versions up to May 4th, capturing net accumulated change rather than every edit, which may mask additions, removals, and reversals. The customisation frame comprised the 621 linked pairs not adopted verbatim (Section~\ref{sec:rq1}), while the evolution frame comprised the 5,978 locally authored skills whose \texttt{SKILL.md} changed (the 8,386 updates mentioned in~\ref{sec:rq3-1} is regarding any co-existing files in a skill).
To suppress outliers, we discarded skills whose churn (added plus removed lines) exceeded the Tukey upper fence of 340 lines, reducing it from 5,978 to 5,367.

Within each setting we stratified the frame into four equal-frequency quartile buckets, by content similarity for the linked pairs, and by churn for the unlinked skills (cut points 8, 34, and 102 lines), so that each stratum contributed an equal number of observations. We sized each stratum for a 90\% confidence level with a $\pm$10\% margin of error under the worst-case Bernoulli variance ($p = 0.5$), applying a finite-population correction per stratum and capping the draw at two pairs per repository to avoid over-representing prolific repositories. This yielded a final coded sample of 188 linked pairs and 256 unlinked skills, evenly distributed across the four buckets.

For the thematic analysis, two authors examined each version pair through a fully rendered \texttt{SKILL.md} diff, with associated commit messages and any changes to co-existing assets surfaced alongside for context. Working first on a common sample of 40 pairs, they extracted key points, the initial step of open coding~\cite{hoda2012self}, each summarising a single discrete change, where one diff could yield several. After agreeing on the granularity to retain, they independently assigned one open code per key point and coded the remaining pairs likewise, resolving disagreements through discussion while cross-verifying each other's codes. Throughout, we compared emerging codes within and across skills to keep them distinct, then merged them into higher-level subthemes and, finally, overarching themes. The complete categorisation was reviewed and agreed upon by the whole team. For each key point we also noted the content type changed against the RQ2 subthemes, on which the two authors aligned.  We did not compute Cohen's $\kappa$ for this field, instead including it within the discussion.  We also reviewed all base codes and classified each as an addition, removal, or modification, which is reported in discussion.


To assess whether the customisation and evolution settings differ in the kinds of changes they accrue, we complemented the descriptive frequencies with a statistical comparison. For each change-type subtheme and theme (Table~\ref{tab:rq3}), we compared the proportion of customisation against evolution skills, using a two-sided Fisher's exact test~\cite{upton1992fisher} with Benjamini-Hochberg procedure~\cite{benjamini1995controlling} to control false discovery rate, and reporting Cohen's $h$~\cite{cohen1988power} effect size. We note that these tests do not account for residual clustering of skills within repositories or dependence among the multiple theme/subtheme labels. As this is an exploratory comparison over an open-coded taxonomy, we read differences as patterns within our coded sample rather than population-level effects, and report non-significant contrasts descriptively.
\section{Skill categories on the SWEBOK (RQ1)}
\label{sec:rq1}

\begin{figure}[t]  
 \centering
\includegraphics[width=0.8\linewidth]{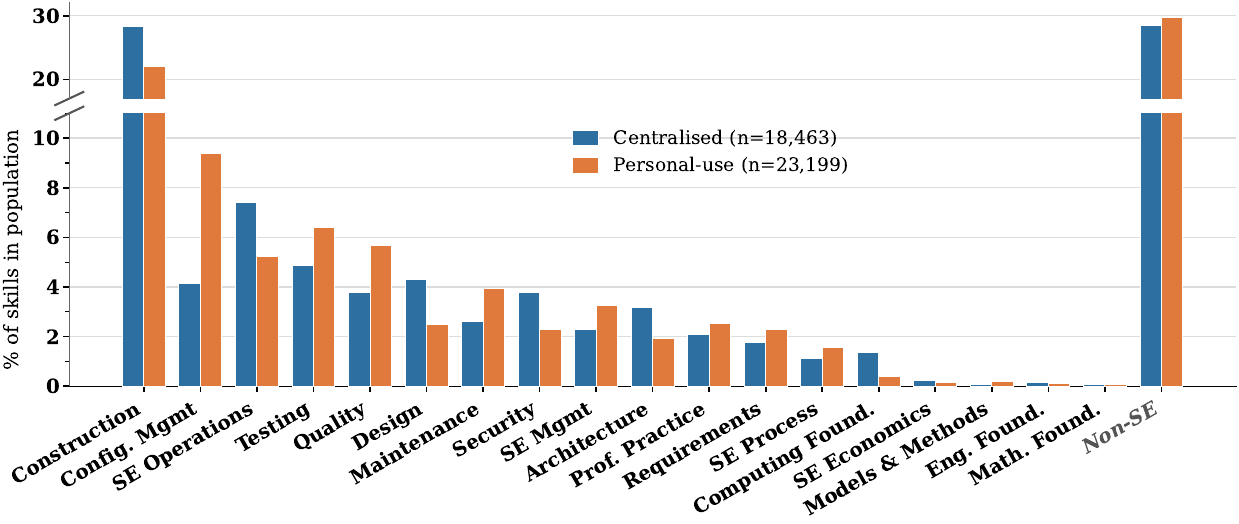}
  \caption{SWEBOK KA distribution of agent skills}
  \label{fig:rq1-ka}
\end{figure}

Applying the SWEBOK-KA classification pipeline (Section~\ref{sec:rq1-method}) to both centralised and personal-use skills, Figure~\ref{fig:rq1-ka} reports their distribution across the 18 SWEBOK KAs alongside a supplementary Non-SE category. Software Construction dominates both populations, comprising 28.3\% of centralised and 22.1\% of personal-use skills. This is followed by Software Engineering Operations ($\approx 5\%$ to $7\%$), Software Testing ($\approx 5\%$ to $6\%$), and Software Configuration Management ($\approx 4\%$ to $9\%$). The remaining 14 KAs form a long tail. Overall, SE skills concentrate on concrete code authoring and execution, whereas domains concerning planning, management, and professional practices remain comparatively under-represented.


The centralised and personal-use skills differ in their distribution across SWEBOK KAs. Centralised skills lean towards broadly reusable, general-purpose coding capabilities. Software Construction is 6.2 percentage points higher in the registry, alongside Software Design (4.3\% vs.\ 2.5\%), Software Architecture (3.2\% vs.\ 1.9\%), and Software Security (3.8\% vs.\ 2.3\%). Conversely, personal-use skills are higher in project-specific lifecycle activities, with Software Configuration Management exhibiting the sharpest divergence (9.4\% vs.\ 4.1\%), followed by Software Testing (6.4\% vs.\ 4.9\%), Software Quality (5.7\% vs.\ 3.8\%), and Software Maintenance (3.9\% vs.\ 2.6\%). Beyond these, a substantial share of skills in both populations (28.5\% vs.\ 29.8\%) do not map to any SWEBOK KA and are categorised as Non-SE skills. Examples such as \textit{babysor/speak} (speech synthesis) indicate usage beyond SE tasks. As this study focuses on SE-related skills, Non-SE skills are excluded from subsequent analyses. After filtering, the dataset comprises 13,206 centralised SE skills across 1,698 repositories and 16,244 personal-use SE skills across 4,656 repositories. Restricting linked pairs to SE skills yields 2,462 reuse connections, covering 1,169 unique centralised SE skills and 2,462 unique personal-use SE skills.  Of these 2,462 pairs, 621 were not adopted verbatim with similarity below 0.99.

\section{Characteristics of AI agent skills (RQ2)}
\subsection{How AI Agent Skills Are Written and Packaged (RQ2.1)}
\label{sec:rq2-1}

\begin{figure}
    \centering
    \includegraphics[width=0.6\linewidth]{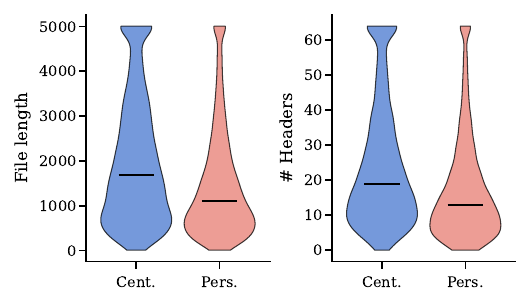}
    \caption{{\texttt{SKILL.md} token length and header count}}
    \label{fig:skillmd-structure}
\end{figure}

The \texttt{SKILL.md} document contains the primary instructions a coding agent reads at invocation. Figure~\ref{fig:skillmd-structure} compares two structural properties: token length and the number of Markdown headers. Centralised skills are systematically longer and more sectioned than their personal-use counterparts. The median \texttt{SKILL.md} spans 1,678 tokens against 1,114, with 19 headers vs.\ 13. Both differences are significant under a Mann-Whitney U test~\cite{mann1947test} ($p<0.001$) with small effect sizes (Cliff's $\delta = 0.23$ and $0.25$~\cite{cliff1993dominance}), suggesting skills intended for shared use carry richer content and finer segmentation. Header Levels 1 and 2 appear in over 90\% of skills, while Level 3 reaches 79.8\% (centralised) and 67.9\% (personal-use). Deeper levels are far rarer, making the hierarchy generally flatter than human-targeted README files~\cite{gao2025adapting}.

Beyond the body, a skill's frontmatter and co-existing assets reveal its packaging discipline. Against the Agent Skills specification (Table~\ref{tab:conformance}), both populations conform well on mandatory clauses ($\geq 99\%$ for \texttt{SKILL.md} presence, frontmatter validity, and \texttt{description} typing and length), but a gap emerges elsewhere. Personal-use skills less often match the parent directory name (91.23\% vs.\ 97.65\%) and the required \texttt{name} character format. Spec-prescribed optional fields (\texttt{license}, \texttt{compatibility}, \texttt{metadata}, \texttt{allowed-tools}) appear in only 3--16\% of skills, reflecting looser discipline outside curated marketplaces. The gap extends to assets. Specifically, \texttt{references/} is the most common optional subdirectory (31.0\% vs.\ 17.0\%), followed by \texttt{scripts/} and \texttt{assets/}, with centralised skills consistently using each more often. 

  \begin{table}[t]
    \centering
    \caption{Conformance and feature presence}
    \label{tab:conformance}
    \small
    \renewcommand{\arraystretch}{0.9}
    \setlength{\tabcolsep}{4pt}
    \resizebox{0.8\columnwidth}{!}{%
    \begin{tabular}{@{}lrr@{}}
      \toprule
      Provision & Cent. (\%) & Pers. (\%) \\
      \midrule
      \multicolumn{3}{@{}l}{\textit{Mandatory provisions: conformance rate}} \\
      SKILL.md exists at skill root              & 100.00 & 100.00 \\
      File opens with valid YAML frontmatter     & 99.80  & 99.99  \\
      \texttt{name} present and non-empty        & 99.79  & 98.37  \\
      \texttt{name} matches required char.\ format & 96.35 & 95.04 \\
      \texttt{name} equals parent directory name & 97.65  & 91.23  \\
      \texttt{description} present, non-empty, string & 99.80 & 99.95 \\
      \texttt{description} $\leq$ 1024 characters & 99.64  & 99.86  \\
      \midrule
      \multicolumn{3}{@{}l}{\textit{Optional fields and directories: presence rate}} \\
      \texttt{license} declared                  & 16.06  & 9.08   \\
      \texttt{compatibility} declared            & 3.98   & 3.18   \\
      \texttt{metadata} declared                 & 13.08  & 10.37  \\
      \texttt{allowed-tools} declared            & 14.95  & 12.47  \\
      \texttt{scripts/} present                  & 14.42  & 8.37   \\
      \texttt{references/} present               & 30.99  & 17.00  \\
      \texttt{assets/} present                   & 4.76   & 1.87   \\
      \midrule
      \multicolumn{3}{@{}l}{\textit{Body and reference recommendations: conformance rate}} \\
      Body $\leq$ 500 lines                      & 84.54  & 93.67  \\
      Body $\leq$ 5{,}000 whitespace tokens      & 99.06  & 99.42  \\
      Markdown links relative, $\leq$ 1 dir.\ deep & 94.08 & 95.36 \\
      \bottomrule
    \end{tabular}}
  \end{table}

\subsection{Content of AI agent skills (RQ2.2)}

Table~\ref{tab:rq2} presents the results for skill content. In total, we derived 69 codes, grouped into 17 subthemes and consolidated into six themes (Table~\ref{tab:rq2}), each detailed below. Because the analysed documents are sampled evenly across KAs, the frequencies reported here and in Table~\ref{tab:rq2} describe this stratified sample rather than corpus-wide prevalence, and are best read as evidence of how consistently each theme recurs across KAs.

\begin{table}[t]
  \centering
  \small
  \setlength{\tabcolsep}{3pt}
  \renewcommand{\arraystretch}{0.9}
  \caption{Categories of skill content.}
  \label{tab:rq2}
  \resizebox{0.8\columnwidth}{!}{%
  \begin{tabularx}{\linewidth}{@{}p{1.6cm}Xrr@{}}
  \toprule
  Theme & Subtheme & \#codes & Freq. \\
  \midrule
  \multirow{2}{1.6cm}{\raggedright \textbf{Scoping and orchestrating}}
    & Defining scope and activation & 3 & 100\% \\
    & Sequencing and branching workflow & 8 & 68\% \\
  \cmidrule(l){2-4}
    & \textbf{Total} & \textbf{11} & \textbf{100\%} \\
  \midrule
  \multirow{5}{1.6cm}{\raggedright \textbf{Running execution lifecycle}}
    & Preparing the environment & 4 & 25\% \\
    & Documenting tool and script usage & 7 & 76\% \\
    & Gathering input information & 3 & 22\% \\
    & Shaping the output artefact & 4 & 50\% \\
    & Providing reusable code and config & 3 & 32\% \\
  \cmidrule(l){2-4}
    & \textbf{Total} & \textbf{21} & \textbf{89\%} \\
  \midrule
  \multirow{3}{1.6cm}{\raggedright \textbf{Ensuring output quality}}
    & Monitoring and inspecting state & 2 & 23\% \\
    & Verifying and evaluating output & 5 & 33\% \\
    & Handling failures and recovering & 3 & 50\% \\
  \cmidrule(l){2-4}
    & \textbf{Total} & \textbf{10} & \textbf{69\%} \\
  \midrule
  \multirow{2}{1.6cm}{\raggedright \textbf{Governing agent conduct}}
    & Constraining and prohibiting & 4 & 45\% \\
    & Recommending best practices & 5 & 77\% \\
  \cmidrule(l){2-4}
    & \textbf{Total} & \textbf{9} & \textbf{81\%} \\
  \midrule
  \multirow{3}{1.6cm}{\raggedright \textbf{Grounding domain knowledge}}
    & Explaining concepts and mechanisms & 2 & 36\% \\
    & Enumerating domain facts and defaults & 2 & 38\% \\
    & Pointing to references and resources & 5 & 72\% \\
  \cmidrule(l){2-4}
    & \textbf{Total} & \textbf{9} & \textbf{85\%} \\
  \midrule
  \multirow{2}{1.6cm}{\RaggedRight \textbf{User/Agent Coordination}}
    & Interacting with the user & 6 & 63\% \\
    & Delegating and handing off work & 3 & 24\% \\
  \cmidrule(l){2-4}
    & \textbf{Total} & \textbf{9} & \textbf{68\%} \\
  \bottomrule
  \end{tabularx}}
\end{table}

\subsubsection{Scoping and orchestrating the task}
This theme captures the scenarios a skill handles and its prescribed execution workflow. Present in all 180 skills, \textit{Defining scope and activation} is the most frequent subtheme, detailing a skill's purpose and triggering conditions. These are typically placed in the YAML frontmatter, preloaded for progressive disclosure, though often restated in the body. To refine activation, some skills also delineate out-of-scope scenarios. A documentation skill, for example, instructs, \textit{``Do NOT use for code comments, inline JSDoc...''}.
\textit{Sequencing and branching the workflow} appears in 68\% of skills, detailing how tasks are executed. Workflows are often end-to-end numbered sequences, though many add conditional logic to guide behaviour across scenarios, not a fixed path. A smaller subset specifies post-execution actions or exit criteria. A flaky-test skill, for example, enforces a checklist directing the agent to identify the root cause, apply a fix, and verify across runs before concluding.


\subsubsection{Running the execution lifecycle}
This theme captures the core knowledge an agent requires to execute a task. The lifecycle typically begins with \textit{preparing the environment}, a subtheme present in 25\% of skills. While some tasks assume pre-existing dependencies like Git, the dominant pattern involves explicit instructions for installing required packages and configuring authentication keys. Certain skills even extend to hardware requirements. For instance, a reinforcement-learning skill explicitly recommends a high-performance GPU setup (\textit{``NVIDIA A100/H100 recommended''}). Following setup, 22\% of skills define the required input format or its precise location in the project repository to ensure accurate execution.


To execute tasks, \textit{documenting tool and script usage} details operational materials, appearing in 76\% of skills. These include bundled scripts or external third-party tools, alongside parameter options and expected output formats. For example, a code-tracing skill invokes ``\textit{grepai trace callees}'' and explains its structured output so the agent can interpret the returned file locations. In contrast, \textit{providing reusable code and config} supplies source material the agent embeds directly into the project rather than executing. For instance, a ``\textit{writing-dockerfiles}'' skill provides a ready-made ``\textit{.dockerignore}'' for verbatim insertion. The final subtheme, \textit{shaping the output artefact}, appears in 50\% of skills. These instructions dictate the artefact's content, format, and file structure, often supplying a reusable template. For example, a planning skill provides a Markdown \textit{Plan Template} prescribing mandatory sections such as \textit{Overview}, \textit{Prerequisites}, and \textit{Dependency Graph}.

\subsubsection{Ensuring output quality} This theme covers instructions for the agent to verify its work and recover from errors. \textit{Monitoring and inspecting state} (23\%) directs the agent to execute diagnostic commands to observe ongoing execution. \textit{Verifying and evaluating output} (33\%) involves assessing the final generated artefact, often operationalised through multi-dimensional checklists or severity rubrics to prioritise fixes. For example, a documentation skill outlines a ``\textit{Documentation Health Check}'' spanning \textit{Structure}, \textit{Cohesion}, \textit{Audience}, and \textit{Navigation}. \textit{Handling failures and recovering} (50\%) warns the agent of common pitfalls and provides recovery instructions. Skills typically enumerate failure modes alongside causal explanations. For instance, a Kafka skill flags a ``\textit{Large Messages}'' anti-pattern, noting that ``\textit{Kafka is optimised for small messages ($<$ 1MB)}''. Finally, some skills explicitly outline fallback procedures for the agent to execute when a step fails.


\subsubsection{Governing agent conduct} Unlike the concrete operational instructions of the execution lifecycle, this theme captures normative guidance prescribing how the agent should behave. \textit{Constraining and prohibiting} (45\%) establishes negative instructions detailing actions to avoid. These frequently comprise domain-correctness and compliance constraints. For example, the ``\textit{dd-apm}'' (Datadog APM) skill explicitly forbids ``\textit{user\_id}'' tagging. A smaller, notable subset extends beyond technical boundaries to impose ethical conduct rules, such as mandating inclusive language over gendered pronouns.

\textit{Recommending best practices} (77\%) instead establishes positive behavioural guidance, ranging from technical conventions to workflow approaches. These guidelines typically enumerate points, often including rationales to justify the prescribed approach. For example, a task-management skill advises dividing complex work into ``\textit{separate tasks}'' to enable parallel development and simplify tracking.


\subsubsection{Grounding domain knowledge} Unlike prior prescriptive guidance, this descriptive theme supplies background context to ground agent decisions. \textit{Explaining concepts and mechanisms} (36\%) focuses on qualitative understanding and operational logic, clarifying essential terminology and system mechanics. For instance, a version-control skill explains how ``\textit{git worktrees}'' maintain multiple working directories to ensure correct application. In contrast, \textit{Enumerating domain facts and defaults} (38\%) supplies quantitative parameters and categorical look-up tables over conceptual explanation. It provides empirically tuned thresholds, such as a clean-code skill flagging functions exceeding ``\textit{100 lines}'' or taking ``\textit{more than 3 parameters}''. Finally, \textit{Pointing to references and resources} (72\%) omits inline knowledge, instead directing the agent to external documentation and bundled assets.

\subsubsection{User/Agent Coordination} The final theme shifts from internal reasoning to external engagement with humans, agents, or tools. \textit{Interacting with the user} (63\%) shapes human-facing behaviour and the broader interaction contract. A common mechanism involves gating consequential actions behind user consent, such as a project-management skill requiring confirmation before merging pull requests or deleting branches. \textit{Delegating and handing off work} (24\%) governs tasks the agent offloads. These skills dictate routing work to specialised agents, transferring data, or escalating to human experts. For example, an alignment skill acts as a ``\textit{front-door alignment layer}'', initially invoking a companion requirement-shaping skill before assuming execution.

Finally, we examine how this thematic content varies across the SWEBOK KAs. A typical skill covers four core themes (scoping the task, running its execution, governing conduct, and grounding domain knowledge) regardless of KA, whereas ensuring output quality and user/agent coordination depend heavily on the area. Only half the skills in Software Design, Configuration Management, Security, and Engineering Foundations address output quality, and coordination ranges from 20\% in Security to 100\% in Requirements, reflecting how human-facing the work is. Subtheme presence varies more sharply. Shaping the output artefact differentiates most (0\% in Computing Foundations to 90\% in Configuration Management), a contrast hidden beneath the near-universal execution lifecycle, while user interaction varies similarly. These patterns reflect the underlying workload: areas producing structured deliverables prioritise shaping the output artefact, human-facing areas emphasise user interaction, and operational domains focus on environment preparation and failure recovery.

\begin{table*}[t]
  \centering
  \small
  \setlength{\tabcolsep}{5pt}
  \renewcommand{\arraystretch}{0.9}
  \caption{Change-type themes and subthemes in skill customisation and evolution.  ($^{*}q<0.05$, $^{**}q<0.01$)}
  \label{tab:rq3}
  \resizebox{0.7\textwidth}{!}{%
  \begin{tabular}{@{}p{3.4cm}lrrrr@{}}
  \toprule
  Theme & Subtheme & \#codes & Customisation & Evolution & Cohen's $h$ \\
  \midrule
  \multirow{4}{*}{\parbox{3.4cm}{\raggedright \textbf{Configuring skill metadata}}} & Supplying or expanding activation metadata & 3 & 36 (19.1\%) & 85 (33.3\%) & 0.33$^{**}$ \\
   & Pruning or repairing activation metadata & 3 & 26 (13.8\%) & 19 (7.5\%) & 0.21 \\
   & Controlling invocation and identity & 2 & 9 (4.8\%) & 12 (4.7\%) & 0.00 \\
   & Managing provenance and distribution metadata & 3 & 33 (17.6\%) & 15 (5.9\%) & 0.37$^{**}$ \\
  \cmidrule(l){2-6}
   & \textbf{Total} & \textbf{11} & \textbf{76 (40.4\%)} & \textbf{110 (43.1\%)} & \textbf{0.05} \\
  \midrule
  \multirow{6}{*}{\parbox{3.4cm}{\raggedright \textbf{Reworking operational specifications}}} & Reshaping the prescribed workflow & 4 & 38 (20.2\%) & 98 (38.4\%) & 0.40$^{**}$ \\
   & Adjusting environment setup and permissions & 4 & 24 (12.8\%) & 26 (10.2\%) & 0.08 \\
   & Revising tool usage guidance & 3 & 35 (18.6\%) & 61 (23.9\%) & 0.13 \\
   & Adjusting outputs and worked examples & 3 & 29 (15.4\%) & 70 (27.5\%) & 0.30$^{**}$ \\
   & Adjusting delegation to other skills or agents & 3 & 13 (6.9\%) & 10 (3.9\%) & 0.13 \\
   & Re-scoping the task and context fit & 3 & 28 (14.9\%) & 17 (6.7\%) & 0.27$^{*}$ \\
  \cmidrule(l){2-6}
   & \textbf{Total} & \textbf{20} & \textbf{107 (56.9\%)} & \textbf{157 (61.6\%)} & \textbf{0.09} \\
  \midrule
  \multirow{2}{*}{\parbox{3.4cm}{\raggedright \textbf{Recalibrating behavioural constraints}}} & Adjusting hard enforcement and constraints & 4 & 31 (16.5\%) & 71 (27.8\%) & 0.28$^{*}$ \\
   & Adjusting softer guidance and recommendations & 5 & 41 (21.8\%) & 38 (14.9\%) & 0.18 \\
  \cmidrule(l){2-6}
   & \textbf{Total} & \textbf{9} & \textbf{58 (30.9\%)} & \textbf{101 (39.6\%)} & \textbf{0.18} \\
  \midrule
  \multirow{3}{*}{\parbox{3.4cm}{\raggedright \textbf{Adapting knowledge and resources}}} & Revising inline knowledge content & 3 & 37 (19.7\%) & 83 (32.5\%) & 0.29$^{**}$ \\
   & Rewiring reference pointers & 3 & 35 (18.6\%) & 57 (22.4\%) & 0.09 \\
   & Managing bundled assets & 4 & 53 (28.2\%) & 53 (20.8\%) & 0.17 \\
  \cmidrule(l){2-6}
   & \textbf{Total} & \textbf{10} & \textbf{87 (46.3\%)} & \textbf{128 (50.2\%)} & \textbf{0.08} \\
  \midrule
  \multirow{3}{*}{\parbox{3.4cm}{\raggedright \textbf{Maintaining skill compatibility}}} & Refreshing currency with upstream releases & 2 & 17 (9.0\%) & 10 (3.9\%) & 0.21 \\
   & Adapting to renames and tooling substitutions & 2 & 10 (5.3\%) & 62 (24.3\%) & 0.57$^{**}$ \\
   & Repackaging and porting to the host platform & 2 & 23 (12.2\%) & 0 (0.0\%) & 0.71$^{**}$ \\
  \cmidrule(l){2-6}
   & \textbf{Total} & \textbf{6} & \textbf{50 (26.6\%)} & \textbf{68 (26.7\%)} & \textbf{0.00} \\
  \midrule
  \multirow{2}{*}{\parbox{3.4cm}{\raggedright \textbf{Polishing presentation and wording}}} & Reformatting layout and structure & 2 & 92 (48.9\%) & 52 (20.4\%) & 0.61$^{**}$ \\
   & Refining wording and grammar & 2 & 26 (13.8\%) & 41 (16.1\%) & 0.06 \\
  \cmidrule(l){2-6}
   & \textbf{Total} & \textbf{4} & \textbf{105 (55.9\%)} & \textbf{78 (30.6\%)} & \textbf{0.52$^{**}$} \\
  \bottomrule
  \end{tabular}%
  }
\end{table*}

\section{Customisation and evolution of skills (RQ3)}

\subsection{Modification at adoption (RQ3.1)}
\label{sec:rq3-1}


We first quantify how far a skill diverges from the registry once it enters a repository. Recall from Section~\ref{sec:rq1} that, after restricting both populations to SE skills, our linkage recovers 2,462 \emph{reused} skills matching a centralised source and 13,782 \emph{locally authored} skills with no such match. Modification at adoption is not prevalent. Of the 2,462 reused skills, 1,841 were adopted near-verbatim, reproducing their source at a body similarity $\geq$ 0.99, leaving only 621 altered on entry.

To characterise how actively skills are maintained, we examined the per-skill commit history of all 16,244 SE skills in the personal-use repositories (2,462 reused and 13,782 locally authored) over a one-year window up to May 4th. We treat a skill as updated if its files received at least one commit beyond the one that introduced it. Overall, 9,553 skills (58.8\%) were updated at least once. Maintenance is more common among locally authored skills. 8,386 of 13,782 (60.8\%) were updated, typically modestly (a median of 2 follow-up commits).

Reused skills are updated less often. Only 1,167 of 2,462 (47.4\%) received any local change after adoption. A more telling pattern emerges when we contrast a reused skill's local activity with the continued evolution of its upstream source. Among the 1,295 reused skills never locally updated, 40.2\% had their centralised counterpart updated one or more times after the adoption date, changes the local copy never received. Among reused skills that were locally updated and whose upstream also changed, 62.3\% (429 of 689) diverged independently rather than re-incorporating the upstream change, comparing local and upstream versions at the first post-adoption upstream update against a 0.99 body-similarity threshold. In other words, adoption largely behaves as a one-time copy that developers rarely synchronise with upstream revisions. This section establishes \emph{how much} skills change. The edit patterns are discussed in the next subsection alongside evolution.

\subsection{The Nature of Skill Changes (RQ3.2)}
\label{sec:rq3-changes}
Our open coding yielded 60 codes across six themes capturing how developers modify skills. Table~\ref{tab:rq3} reports these frequencies, separating customisation in reuse contexts from evolution. Percentages indicate the proportion of analysed skills undergoing each update. Totals can exceed 100\% since individual skills often experience multiple operation types.

\subsubsection{Configuring skill metadata}
Occurring predominantly within agent-preloaded frontmatter, these edits occur at comparable overall rates between contexts (40.4\% vs.\ 43.1\%). \textit{Supplying or expanding activation metadata} is significantly more frequent during evolution, accommodating new triggers. For example, the ``\textit{handle-pr-review}'' skill extended its frontmatter for pull-request triage. Conversely, \textit{Managing provenance and distribution metadata} skews significantly towards customisation, where adopters strip central-registry licences and authorship claims. \textit{Pruning or repairing activation metadata} also leans towards customisation to condense descriptions, whereas evolution typically removes invalid fields like ``\textit{invocation: user}''. Finally, \textit{Controlling invocation and identity} shows similar rates across contexts, covering adjustments to skill names and invocation methods.

\subsubsection{Reworking operational specifications}
As the most frequent edit type (56.9\% vs.\ 61.6\%), this theme has three significantly divergent subthemes. \textit{Reshaping the prescribed workflow} roughly doubles in evolution frequency, predominantly by adding extra steps. For example, a quality-assurance skill appended unit and E2E testing to its existing checks. \textit{Adjusting outputs and worked examples} likewise appears significantly more often during evolution. Conversely, \textit{Re-scoping the task and context fit} is significantly more frequent during customisation, aligning with repository specifics. For instance, an ``\textit{android-data-layer}'' skill replaced Hilt with Koin to match the project stack.
Within \textit{Revising tool usage guidance}, customisation typically trims instructions to retain essential basics, whereas evolution tunes commands for reliability. ``\textit{Vulnetix/cli}'', for instance, replaced ``\textit{git branch --show-current}'' with ``\textit{git rev-parse --abbrev-ref HEAD}'' to ensure broader environmental compatibility. Within \textit{Adjusting environment setup and permissions}, customisation typically removes inapplicable installations or substitutes local methods. Finally, \textit{Adjusting delegation to other skills} modifies how skills coordinate with companion skills or subagents.

\subsubsection{Recalibrating behavioural constraints}
This theme adjusts agent conduct constraints. \textit{Adjusting hard enforcement and constraints} occurs significantly more often during evolution, predominantly by appending prohibition rules. For instance, the ``\textit{lurk}'' skill hardened its monitoring by explicitly forbidding memory-store writes via raw shell redirects. Conversely, \textit{Adjusting softer guidance and recommendations} integrates project-specific advisory conventions, such as guidelines for branch naming, rather than strict regulations.


\subsubsection{Adapting knowledge and resources}
Appearing in roughly half of all edits (46.3\% customisation, 50.2\% evolution), this theme updates supporting context. \textit{Revising inline knowledge content} is significantly more frequent during evolution, predominantly appending explanatory rationale as domain knowledge. For example, a chat-streaming skill added a ``Why direct SDK?'' note explaining its approach. Within \textit{Rewiring reference pointers}, developers frequently replace hard-coded content with pointers to a centralised source of truth. In \textit{Managing bundled assets}, evolution adds material while customisation prunes irrelevant files. Adapting a browser skill for a focused X (Twitter) workflow, for instance, removed nine obsolete bundled assets.


\subsubsection{Maintaining skill compatibility}
As the least frequent theme (26.6\%  vs. 26.7\%), two of its subthemes show sharply contrasting profiles. \textit{Repackaging and porting to the host platform} occurs exclusively during customisation, the sharpest divergence in our data ($h{=}0.72$; 12.2\% vs.\ 0.0\%): Adopters convert imported skills to match their platform, typically by rewriting frontmatter or substituting local tools. For instance, adapting the ``\textit{implement-plan}'' skill to one repository required stripping its Claude-specific YAML frontmatter to fit a platform-neutral ``\textit{.agents/skills}'' layout.


Mirroring this contrast, \textit{Adapting to renames and tooling substitutions} is overwhelmingly an evolution activity ($h{=}0.57$; 5.3\% vs.\ 24.3\%), predominantly synchronising skills with project-wide structural changes. A component skill, for instance, updated a referenced configuration file (\textit{app-of-apps/values.yaml} $\rightarrow$ \textit{values-base.yaml}) to reflect an upstream restructure. \textit{Refreshing currency with upstream releases} forms a minor category across both contexts, focusing on version bumps or migration from deprecated APIs.


\subsubsection{Polishing presentation and wording}
This theme is significantly more frequent during customisation ($h{=}0.52$; 55.9\% vs.\ 30.6\%). Rather than altering behaviour, these edits refine presentation by adjusting Markdown, tightening prose, and correcting typos. The skew is driven almost entirely by \textit{Reformatting layout and structure} ($h{=}0.61$), the most frequent customisation edit (48.9\%). \textit{Refining wording and grammar}, meanwhile, remains balanced across contexts.

\section{Discussion and Implications}

\subsection{Discussion}

First, RQ1 reveals a distributional difference between centralised and personal-use skills across KAs (Figure~\ref{fig:rq1-ka}), which may reflect a supply-and-demand dynamic. Centralised skills emphasise broadly reusable, general-purpose coding capabilities such as Software Construction and Design, whereas personal-use skills are higher in project-specific lifecycle activities such as Software Configuration Management and Testing. One interpretation is that the registry supplies general coding capabilities, leaving developers to author local skills for project-specific configuration, testing, and maintenance. Tied closely to individual codebases, these offer limited external utility, which may explain why they remain local rather than entering the shared marketplace. This is not evidenced by adoption patterns and may be open to other interpretations.

Second, each base-level code is a key point we extracted for one part of a skill update, and records a direction: ``add'', ``remove'', or ``modify''. Aggregating these reveals a directional asymmetry in how skills change. Excluding modifications, additions outnumber removals by 2.7 to 1 (586 vs.\ 221). Customisation is relatively balanced (1.1:1), adopters removing almost as often as they add, whereas evolution skews heavily towards additions (6.1:1). Maintenance is thus predominantly additive, accruing instructions far more than pruning them, echoing documentation maintenance~\cite{gao2025adapting}. Developers mainly add quality-verification steps, constraints, and best-practice recommendations, while the only frequently deleted content is licences and authorship claims from centralised skills. 

Third, our coding notes on the content changed on a unit of skill updates reveal that revisions are highly uneven across a skill, with customisation and evolution diverging in focus. Two regularities span both settings. First, the instructions that shape a skill's user interaction, runtime monitoring, and failure handling are rarely updated.
Interacting with the user (1.6\% of customisation and 1.2\% of evolution diffs), runtime-state monitoring (0.5\%/0.0\%), and handling failures (5.9\%/7.1\%) rank lowest, so adopters and maintainers inherit these rules almost verbatim. Second, defining scope and activation and documenting tool and script usage rank near the top in both contexts (38.8\%/42.0\% and 30.3\%/36.1\%), so material connecting a skill to specific tasks and toolchains is routinely rewritten. Elsewhere the contexts diverge. Customisation re-grounds a skill to its host project. Specifically, pointing to references and resources and preparing the environment are revised more than in evolution (34.6\%/30.2\% and 16.0\%/10.2\%), as adopters repoint paths and adapt to local infrastructure. Evolution instead expands the core procedure, revising sequencing and branching the workflow, shaping the output artefact, and enumerating domain facts far more often than customisation (23.4\%/38.4\%, 9.0\%/16.1\%, and 3.7\%/12.9\%).

\subsection{Implications}

\textbf{Skill users and maintainers.} In the reuse paradigm, the boundary between users and maintainers blurs as adopters assume responsibility for keeping skills current, yet over half (53\%) receive no updates after adoption. Our findings motivate several recommendations. First, elements reflecting the local development context are the most susceptible to change, so users should configure upstream-change alerts, periodically validate referenced paths and versions, and test bundled scripts as workflows evolve. Like standard documentation, a \texttt{SKILL.md} can become outdated or unsynced without active maintenance~\cite{gao2025adapting}, with more severe consequences than for a human reading stale instructions~\cite{mirhosseini2020docable}. Second, users should monitor agent activation. Because vague criteria can prevent invocation, adopters enumerate the exact scenarios in which a skill should trigger, echoing findings that concise descriptions with specific use cases improve agent tool use~\cite{yuan-etal-2025-easytool}. Finally, the most stable components are those shaping runtime interaction with users, progress monitoring, and failure recovery. These form a behavioural contract that travels with a skill across reuse and maintenance. Authors should therefore craft these rules carefully at creation, since downstream users rarely revise them, leaving underlying issues hard to surface.

\textbf{Registries and platforms.} First, platforms should promote agent-neutral skill designs. We frequently observe skills tailored to a single agent, such as Claude, whose subsequent maintenance largely involves porting instructions to other models like Codex. Platform-agnostic formats would mitigate this redundant effort. Second, registries could adopt prefilled templates to guide how knowledge is specified and structured within a skill. As our qualitative analysis showed, four backbone themes constitute the majority of a skill's content while other elements remain highly variable, so templates with defined mandatory and optional fields would simplify initial construction and ensure baseline quality, as how structured README templates aid project documentation~\cite{gao2025adapting}.

\textbf{Researchers.} First, domain-specific knowledge in skills undergoes heavy modification during evolution, predominantly additions (Table~\ref{tab:rq3}). Such knowledge, typically framework-specific rationales and caveats, is re-authored within individual skills as unstructured text. As SWEBOK formalises software engineering knowledge, structured information across technical stacks could be consolidated into centralised, reusable resources, aided by extraction tools such as those mining critical information from API documentation~\cite{li2018improving} or caveats from Stack Overflow~\cite{treude2016augmenting}. Second, researchers should better understand skill activation and propose interactive tracing methods for developers. Frontmatter descriptions act as preloaded text determining invocation, and we observe frequent updates to fine-tune triggering. Since concise instructions with specific scenarios enhance LLM tool use~\cite{yuan-etal-2025-easytool} and skills are invoked implicitly, robust tracing would help developers calibrate activation criteria. Recent work benchmarks skill usage~\cite{li2026skillsbench} and automatically evolves skills to reduce ``skill technical debt''~\cite{pu2026skillops}. Complementing these with continuous monitoring throughout development could further help developers improve both skills and their instructions. Finally, skills uniquely combine non-deterministic natural-language instructions with deterministic bundled scripts. Prose provides exploratory freedom but may yield non-deterministic behaviour~\cite{ouyang2025empirical}, whereas bundled scripts let agents execute deterministic commands, offering predictability and testability at the cost of prose-afforded flexibility~\cite{tam2024let}. Because this balance likely depends on a skill's nature and KA, investigating when and why maintainers trade off these dimensions is a promising direction.

\section{Threats to Validity}
\textbf{Construct Validity}: The first construct involves our use of SWEBOK KAs to capture the software engineering domain a skill covers. Although SWEBOK enumerates the discipline's established KAs, these are not mutually exclusive, so a skill may touch several. For simplicity, we force each skill into a single KA, assigning only its primary one where it spans several. The second construct is our reliance on the frontmatter name field with a content-similarity distance to recover linkage between centralised and personal-use skills. This misses skills that are heavily edited or renamed, and assumes reuse runs from centralised to personal-use, not the reverse. The final construct concerns the customisation and evolution patterns, whose contrast combines two designs: an immediate registry-to-local adoption diff vs. an initial-to-final endpoint diff. The latter captures net change rather than every action, hiding intermediate additions, removals, and reversals.


\textbf{Internal Validity}: The first concerns the reliability of our qualitative analysis; two authors coded jointly and discussed disagreements until reaching consensus. Second, our linking of personal-use to centralised skills may undercount connections, yielding a conservative estimate of single adoption without synchronisation. Some skills we treat as
  locally authored may thus be undetected reuse from elsewhere, so the evolution arm is better described as unlinked than authored from scratch. Third, our 20-install threshold excludes less-downloaded skills, missing links and customisation in less popular repositories. Finally, KA labels rely on an LLM, sensitive to prompt, decoding settings, model
  version, and reruns; we fixed temperature at 0 for reproducibility. Our gold standard was stratified by Qwen's predicted labels, validating category-level correctness but not corpus-level accuracy or Non-SE prevalence, so reported KA proportions should be read as classifier-based estimates with possible labelling error.


\textbf{External Validity}: The first threat is sampling bias. We collect from GitHub repositories that publish their skills, whereas many projects are not open source or do not publish these configurations. We do not claim generalisability to those settings, but use this subset as a proxy. Relatedly, capping repositories at Tukey's fence removes aggregators but may exclude genuine projects that maintain in-house skill collections, biasing the personal-use sample towards smaller adopters.
Second, as skills are an emerging artefact, our study captures only a snapshot. Specifically, it covers skill content to 4 March and repository history to 4 May 2026. These practices will likely mature as the ecosystem evolves. Finally, we focus on a single centralised registry, \texttt{skills.sh}. Although alternative registries such as SkillMP draw on similar sources, they are a less representative proxy for downstream usage, relying on indirect signals such as repository stars. We therefore consider \texttt{skills.sh} a more reliable basis for studying reuse and expect our findings to generalise to broader reuse patterns.

\section{Conclusion}
We presented the first large-scale empirical study of AI agent skills as engineered, reused, and maintained software artefacts. We mined 18,463 skills from the \texttt{skills.sh} registry and 23,199 personal-use skills across 5,876 GitHub repositories, linked by 3,709 recovered reuse connections. Mapping skills onto the SWEBOK KAs, we found Software Construction dominant amid a long tail. Our thematic analysis of skill content yielded six categories.
We then qualitatively coded updates on 444 \texttt{SKILL.md} files under customisation and evolution into six edit categories, finding that reworking operational specifications and adapting knowledge and resources are the most frequently edited. Customisation re-grounds a skill to its host project, whereas evolution adds new inline domain knowledge. Edits skew heavily towards additions over removals, so locally maintained skills tend to grow over time. These findings characterise how skills are written, reused, and maintained, and yield actionable implications for skill maintainers, registries, and researchers.

\section{Replication Package}
\label{sec:replication}
The replication package for this study is available at \url{https://zenodo.org/records/21032973}.

\ifCLASSOPTIONcaptionsoff
  \newpage
\fi

\footnotesize{
\bibliographystyle{IEEEtranN}
\bibliography{reference}
}

\end{document}